\documentclass{ws-procs9x6}
\def\etal{{\it et al.}}

\newcommand{\beq}{\begin{equation}}
\newcommand{\eeq}{\end{equation}}
\newcommand{\bea}{\begin{eqnarray}}
\newcommand{\eea}{\end{eqnarray}}
\newcommand{\bsub}{\begin{subequations}}
\newcommand{\esub}{\end{subequations} \noindent}

\renewcommand{\theequation}{\thesection.\arabic{equation}}

\def\PRD#1#2#3{Phys. Rev. {\bf D#1} (#2) #3}

\def\NPB#1#2#3{Nucl. Phys. {\bf B#1} (#2) #3}
\def\PTP#1#2#3{Prog. Theor. Phys. {\bf #1} (#2) #3}

\def\PLB#1#2#3{Phys. Lett. {\bf B#1} (#2) #3}
\def\PRL#1#2#3{Phys. Rev. Lett. {\bf #1} (#2) #3}
\def\PR#1#2#3{Phys. Rep. {\bf #1} (#2) #3}
\begin{document}
\begin{center}
    \vspace*{0.5cm}
    {\large\bf NEUTRINO BI-LARGE MIXINGS AND  FAMILY 
\footnote{
\uppercase{T}alk presented at \uppercase{NOON}2003 held at 
\uppercase{K}anazawa, \uppercase{F}eb.~10 -14, 2003.
}
}\\
\vspace{0.2cm}
    {\large
      Masako {\sc Bando}\footnote{E-mail address:
        bando@aichi-u.ac.jp} 
      and Midori {\sc Obara}\footnote{E-mail address:
        midori@hep.phys.ocha.ac.jp}}   \\
    \vspace{4mm}
    $^1$ {\it Aichi University, Miyoshi，Aichi 470-0296, Japan} \\[1mm]
    $^2$ {\it Institute of Humanities and Sciences, \\ 
Ochanomizu University, Tokyo 112-8610, Japan}
\end{center}
\abstracts{
After a brief review of quark-lepton relations in grand unified 
theories (GUT), 
we show that the Pati-Salam relation with only one type of Higgs field configuration with 
"four  zero symmetric texture"   can reproduce 
two large neutrino mixings as well as observed mass differences.
This is quite in contrast to the case of $SU(5)$ where bi-large mixings 
essentially come from 
the charged lepton sector with non-symmetric charged lepton mass matrix. 
}
\section{Neutrino Masses and GUT}
Recent results from KamLAND~\cite{KamLAND} 
together with  the neutrino experiments by 
Super-Kamiokande~\cite{kamioka:2002pe,kamioka:2001nj} and 
SNO~\cite{SNO} have confirmed  
neutrino oscillations with two large mixing angles~
\cite{Valle,Fogli,Bahcall,smirnov} 
with the mass squared differences are 
\begin{eqnarray}
0.29 \le  \tan^2 \theta_{12} \le 0.86,\,\, 
&&5.1\times 10^{-5} \le  \Delta m^2_{12} \le 9.7 \times 10^{-3}~\rm{eV}^2,  \\ 
0.83 <  \sin^2 2\theta_{23}, \,\, 
&& 1.4\times 10^{-3} \le  \Delta m^2_{23} \le 6.0 \times 10^{-3}~\rm{eV}^2, 
\label{expangle}
\end{eqnarray}
As we can express the neutrino mixings 
in terms of  MNS matrix~\cite{MNS},  which are 
further divided into two terms, 
$U_l$ and $U_{\nu}$, the unitary matrices which diagonalize 
the $3\times 3$ charged lepton and neutrino mass matrices, 
$M_l$ and $M_{\nu}$;   
\begin{eqnarray}
U_{MNS}&=&U_lU_{\nu}^{\dagger}, \\
U_l^{\dagger}M_lV_l&=&{\rm diag}(m_e,m_{\mu},m_{\tau}), \quad 
U_{\nu}^{\dagger}M_{\nu}U_{\nu}={\rm diag}(m_{\nu_1},m_{\nu_2},m_{\nu_3}), 
\label{MNS}
\end{eqnarray}
in analogous to CKM matrix derived from quark mass matrices, $M_d$ and $M_u$; 
\begin{eqnarray}
U_{CKM}&=&U_uU_{d}^{\dagger}, \\ 
U_u^{\dagger}M_uV_u &=& 
{\rm diag}(m_u,m_c,m_t),  \quad 
U_d^{\dagger}M_dV_d = 
{\rm diag}(m_d,m_s,m_b), 
\label{CKM}
\end{eqnarray}
where $U$ and $V$ are unitary matrix acting on left- and right-handed 
fermions, respectively and ${\rm diag}(m_{\nu_1},m_{\nu_2},m_{\nu_3})$ 
are mass eigenvalues of fermions. 
The observed tiny neutrino masses are  
most naturally explained if tree mass terms are forbidden by some symmetry 
and they come only from the higher dimensional operator (the so-called see-saw mechanism), 
which we adopt in this paper. 
Then the neutrino mass matrix 
$M_{\nu}$ is derived from huge right-handed Majorana masses ($M_R$) and 
the Dirac masses ($M_{\nu_D}$) of EW scale; 
\begin{eqnarray}
M_{\nu}=M_{\nu_D}^T M_R^{-1} M_{\nu_D}. 
\label{seesaw}
\end{eqnarray}
In the SUSY standard model, the  following Yukawa terms, 
which has family indices $3\times 3$,  $Y_{i,j}$, 
connect $SU(2)_L$ ${\bf 2}$, $Q$ and $L$ to ${\bf 1}$ fermions, 
$u_R,d_R,e_R,\nu_R$; 
\begin{equation}
W_Y=Q_L(Y_uu_RH_u+Y_dd_RH_d)+L(Y_le_RH_u+Y_{\nu_D}\nu_R H_d)
+Y_R\nu_R\nu_RH_R. 
\label{yukawa}
\end{equation}
The $SU(2)$ doublet up- and down-type Higgs fields, $H_u,H_d$ 
with VEVs, $v_u,v_d$  give mass matrices 
$\rightarrow M_u=Y_uv_u, M_d=Y_dv_d, M_{\nu_D}=Y_{\nu_D}v_u, M_l=Y_lv_d$
 after the standard symmetry is broken down to 
$SU(3)\times U(1)_{em}$.
The Majorana mass term  (the last term of Eq.~(\ref{yukawa})  
including $SU(2)$ singlet Higgs 
field $H_R$ gives neutrino right-handed mass term, $M_R=Y_Rv_R$ where 
the VEV $v_R$ is expected to be much larger than EW scale. 
\section{Hierarchy Problem}
Before going to the discussion of neutrino masses, we make comments on the hierarchy problems 
whose origin may indicate something to the family structure. 
If one wants to construct unified theory, it is governed by the  
scale $M_P$  which is far higher than electroweak scale. 
We know that the GUT scale, $M_G$ is near below the Planck scale, 
$M_P$ which is usually taken as 
the reduced Planck mass, $1/\sqrt{8\pi G_N}$. 
This huge discrepancy between two scales, $M_P$ and $M_W$ is called 
"strong hierarchy problem". The introduction of supersymmetry (SUSY) 
provides a good solution for solving strong hierarchy. 
The GUT itself needs  several energy scales appearing in 
the steps of GUT  breakings into standard symmetry $G_1\rightarrow G_2 \cdots G_{M}$, 
with each of scales actually expressed as $\lambda^nM_P$. 
There we use the typical 
hierarchical parameter  $\lambda\sim 0.2$ (the Cabibbo mixing angle). 
Especially recent neutrino small masses indicates some intermediate 
scales of order $10^{11-13}$ GeV, which  is roughly equal to 
 $\lambda^8M_P$. Thus we here recognize "mild hierarchy" appearing 
as $\lambda^nM_P$ in  terms of the typical 
hierarchical parameter  $\lambda\sim 0.2$ (the Cabibbo mixing angle). 
The intermediate 
scale $M_R$ is of order $10^{11-13} $ GeV corresponds to $\lambda^8M_P$. 
On the other hand, the hierarchical  fermion masses 
are  also controlled by the same $ \lambda$; the electron mass, 
the smallest Dirac fermion mass is almost  $\lambda^8M_t$, 
in terms of the top quark mass, $m_t=170$ GeV. 
It is interesting that both in  high and law energy regions 
(at $M_P$ and $M_W$ scales) common 
 the power law,  $\lambda^n$ mild hierarchy structure, 
which may give a hint of the origin of family. 
Indeed the simplest example to explain this is 
to introduce the anomalous $U(1)_X$ family quantum number:  
the power structure comes from the Froggatt-Nielsen mechanism according to 
which Yukawa couplings come from higher dimensional operators 
$\lambda^n=(\frac{\theta}{M_P})^n$ with $n$ determined to compensate the 
$U(1)_X$ symmetry by Froggatt-Nielsen field $\theta(X=-1)$. 
This simplest example of family symmetry has a characteristic feature; 
If we assume the strength of all coupling constants of order $1$, 
the Froggatt-Nielsen mechanism produces the power hierarchical coefficients 
of $\lambda^n$ where $n$ is solely determined by the $U(1)$ charges of relevant 
fields and so such power structure is always of factrizable form. 
Within the framework of GUT the $X$ charge is assigned to the each GUT multiplet. 
Also this $U(1)_X$ explains the  "mild  hierarchies" of symmetry breaking scales 
by assigning $X$ charges to Higgs fields. 
We could also introduce more complicated family symmetries beyond abelian case. 
In the following we shall examine how such family quantum numbers can 
be consistent with 
recent neutrino experimental data. 
\section{GUT and Family symmetry}
Now if  we assume some GUT and that 
the family structure is the same for all the members of a multiplet. 
Two types of relations are derived between the unitary matrices 
$U_{u,d,l,\nu}$ 
according to the different kinds of symmetries of the system. 
In the following sections we investigate how the unitary matrices, 
$ U_d, U_u, U_l$ and $U_{\nu_D}$, are mutually related with each other 
if we assume some grand unified gauge theory or some family symmetry.

First, within the standard model, there is no relation between them 
 since the up and down fermions couple to 
different Higgs fields. 
If the hierarchical mass structure comes from 
the Froggatt Nielsen mechanism, then the power factor 
is determined merely from the $X$ charges of relevant left-handed fermions. 
Then, since we have already 
$SU(2)_L$ in standard model, the  $SU(2)$-doublets, $Q$ and $L$ have common charges, 
$q_1,q_2,q_3$ and $l_1,l_2,l_3$, respectively. 
yielding 
\begin{equation}
U_{CKM}\simeq U_d \simeq U_u,\quad   U_{MNS}\simeq U_l \simeq U_{\nu_D}\simeq U_{\nu}, 
\label{smmixing} 
\end{equation} 
with $\simeq$ indicating the same power structure. 
Second, in  the Georgi-Glashow $SU(5)$ symmetry 
($S(3)_c\times SU(2)_L\times U(1)_Y \subset SU(5)$ ), where 
 $(Q_L,e_R,u_R)$ belong to the same multiplet, ${\bf 10}$, while 
 $(L, d_R)$, to ${\bf 5*}$. 
Note that $SU(5)$ GUT never provides new relations for 
 $M_{CKM}$ and $M_{NMS}$, although 
we have a  familiar relation, $M_l \leftrightarrow  M^T_d$, once we fix 
the representations of Higgs field, 
$  M_d \leftrightarrow  M^{T}_l$, 
where $\leftrightarrow$ means that they are mutually related by accompanying some CG 
coefficients according to the representations of coupled Higgs field. 
Thirdly, the Pati-Salam symmetry $SU(4)_{PS}$ combines 
$u_L$ with $\nu_L$ into $F_L$ and $d_L$ with $l_L$ into $F_R$, so we have 
$ U_{CKM} \simeq U_u\simeq U_d  \simeq  U_{MNS} \simeq U_l \simeq U_{\nu}$. 
Thus we see that the Pati-Salam symmetry with $X$-charge power law 
does predict the same hierarchical mixing matrix both for $U_{CKM}$ and $U_{MNS}$. 
Thus in order to reproduce neutrino large mixing angles within Pati-Salam symmetry, 
we should discard 
simple Froggatt Nielsen mechanism and give up factrizable property of power law structure. 
The simplest example of such possibilities is to introduce "zero texture" which has been 
extensively investigated by many authors\cite{symtex}. 
Note that even in such case we have the relation of  mass matrices, 
$ M_{\nu_D} \leftrightarrow  M_u, \quad M_l \leftrightarrow M_d$. 
For larger GUT, $SO(10)$, the situation is essentially the same, so far as they include
the subgroups above mentioned. 
In $SO(10)$ all the fermions of a family may form a single multiplet ${\bf 16}$, 
and so if  the Froggatt Nielsen mechanism works, 
all the components of $\psi({\bf 16})$ do have common 
$X$ charge. So if we take $\psi_i({\bf 16})$ to $i$-th family, 
and since Pati-Salam symmetry dictates small neutrino mixing angles. 
Thus  we must introduce twisted family structure by introducing new fermions $\psi({\bf 10})$ 
and $\psi_i({\bf 16})$ must not be identified to $i$-th family 
(non-parallel family structure), which is most naturally implemented 
in $E_6$ GUT\cite{twist}. 
Leaving the discussion of $SU(5)$ symmetry to Kugo's talk, I here  
show an example of the Pati-Salam symmetry in the next section. 
\section{Pati-Salam Symmetry with Symmetric Four Zero Texture}
The model we introduce here is the following example\cite{obarabando} 
of symmetric "four zero" texture 
which has been extensively investigated 
by many authors\cite{symtex}. 
Under the Pati-Salam symmetry ( we name  this  
"up-road option" because the neutrino large mixing angle 
is related to up-quark mass matrix $M_u$ as mentioned in the previous section). 
We show that  the following 
configuration of the representation of Higgs field for up-quark mass matrix
\begin{eqnarray}
M_u =
\left(
\begin{array}{@{\,}ccc@{}}
 0                 &{\bf 126}           & 0   \\
{\bf 126}           &{\bf 10}            &{\bf 10} \\
 0                 &{\bf 10}            & {\bf 126}
\end{array}\right),\, 
M_R =\left(
\begin{array}{@{\,}ccc@{\,}}
 0          &  rM_R             & 0   \\
 rM_R          &  0            &0 \\
 0                 &0       & M_R
\end{array}\right)\, .    
\label{massmat}
\eea
which may be compared with the Georgi-Jahrscog of down-type  mass matrix. 
Now $M_{\nu_D}$ is obtained  by 
multiplying Clebsch-Gordan coefficient, $1$ or $-3$; 
\begin{eqnarray}
M_u 
\simeq m_t\left(
\begin{array}{@{\,}ccc@{\,}}
0 & a & 0 \\ 
a & b & c \\ 
0 & c & 1 
\end{array}
\right)  
 \, 
\leftrightarrow M_{\nu_D}
=m_t\left(
\begin{array}{@{\,}ccc@{\,}}
0 & -3a & 0 \\ 
-3a & b & c \\ 
0 & c & -3
\end{array} 
\right),\, 
\begin{array}{@{\,}c@{\,}}
a= \frac{\sqrt{m_u m_c}}{m_t},  \\
b=\frac{m_c}{m_t}, \\
c=\sqrt{\frac{m_u}{m_t}}.
\end{array}
\label{patisalamdnu}
\end{eqnarray}
Then $M_{\nu}$ are easily calculated. 
In order to get large mixing angle $\theta_{23}$, the 
of 2-3 element of $M_{\nu}$  should be of the same 
order of magnitude  as 3-3 element, namely, 
$h \simeq \frac{-ac}{3r} \sim 1 \rightarrow
r \sim \sqrt{\frac{m_u^2m_c}{3m_t^3}} \sim10^{-7}$. 
Such kind of enhancement mechanism is 
called "seesaw enhancement"~\cite{Bando:1997ns,tanimoto,smirnov}
With tiny $r$,  $M_{\nu}$ is approximately given, 
\begin{eqnarray}
M_{\nu} 
= \left(
\begin{array}{@{\,}ccc@{\,}}
 0                 &\frac{ 9a^2}{r}           & 0   \\
\frac{9a^2}{r}   &\frac{-6ab}{r} & \frac{-3ac}{r} \\
 0  & \frac{-3ac}{r} & 9
\end{array}
\right) \frac{m_t^2}{m_R}
\equiv  
\left(
\begin{array}{@{\,}ccc@{\,}}
 0                 &\beta           & 0   \\
\beta   &\alpha     & h \\
 0  & h  & 1
\end{array}
\right) \frac{9m_t^2}{m_R} , 
\,\,  
\begin{array}{@{\,}c@{\,}}
h=-ac/3r,  \\
\alpha=2hb/c, \\
\beta= -3ha/c, 
\end{array}
\label{apnmu}
\end{eqnarray}
where $\sin^2 2\theta_{23}$ can be made large when $h\sim \mathcal{O}(1)$. 
However it is non-trivial to reproduce both the experimental bound $\sin^2 2\theta_{12}$ 
and mass ratio $m_{\nu_{12}}/m_{\nu_{23}}$ and 
we should not stay in order-of-magnitude calculation. 
Since all the matrix elements of $M_{\nu}$ are now expressed in terms of up-quark masses 
with the parameter $r$  or $h$, the neutrino masses and mixing angles at GUT scale are 
obtained straightforwardly\cite{obarabando}. 
The obtained formula Eq.~(\ref{apnmu}) at GUT scale is to be compared with the 
neutrino experimental data at GUT scale. Here we estimate the RGE evolution  
of neutrino mass matrix at $M_Z$ by using the approximate formula obtained 
by Haba and Okamura\cite{habaokamura}. 
the transformed expression of  $\alpha, \beta$ and $h$ of Eq.~(\ref{patisalamdnu})
from $M_R$ to $m_Z$ scale are given as, 
\begin{equation}
 \beta \to \frac{1}{(1-\epsilon_e)(1-\epsilon_{\mu})} \beta,  \quad 
 \alpha \to  \frac{1}{(1-\epsilon_{\mu})^2} \alpha,  \quad 
 h \to  \frac{1}{1-\epsilon_{\mu}} h. 
\end{equation}
where the RGE factors  $\epsilon$'s is estimated as at most $0.1$, 
We compare the corrected values of neutrino masses and mixing angles at $M_Z$ scale 
and the experimental data and found that within the  
error
only the small region of $h$ near around 
$1$ is consistent with the observed mass ratio and two mixing angles, for which 
case the final expressions of neutrino masses and mixing angles at GUT scale is,  
\begin{eqnarray}
\tan^2 2\theta_{23} &\simeq& \frac{4}{(1-2xy)^2},\,\,\,                  
\tan^2 2\theta_{12}  \simeq  \frac{72x^2}{(1-2xy)^2},\,\,\,     
\sin \theta_{13} \simeq   \frac{3x}{(3+2xy)},\nonumber \\ 
\frac{m_{\nu_2}}{m_{\nu_3}}&\simeq & \frac{(2xy-1)(1+\sec\theta_{12})}{(3+2xy)}, \quad  
\frac{m_{\nu_1}}{m_{\nu_2}}\simeq  \frac{(1-\sec\theta_{12})}{(1+\sec\theta_{12})}, \nonumber
\end{eqnarray}
with $x=\sqrt{\frac{m_c}{m_t}}, y=\sqrt{\frac{m_c}{m_u}}$. 
We leave the detailed calculations in our full paper\cite{obarabando}. 
and here note that our model predicts not only the order of magnitudes but 
the exact vales of all neutrino masses and mixing angles. This is remarkable and 
enables us to predict $U_{e3} \le 0.1$ irrespectively of CP phase. 
It is remarked that our neutrino mass matrix has been determined with almost uniquely 
determined within error bars of up-quark masses, it can make the prediction of 
leptgenesis calculation once we fix the CP phase, which are now under calculation 
by Obara, Kaneko Tanimoto and Bando\cite{leptogenesis}. 
In conclusion we have seen that the up-road option 
can reproduce the present neutrino experimental data very well. 
However also down-road option may be also 
worthwhile to be investigated\cite{twist}, in which case the Nature 
may show "twisted family structure".  On the contrary 
in the case of up-road option it requires "parallel family structure". 

We thank to M.~Tanimoto,  A.~Sugamoto and T.~Kugo and 
S.~Kaneko
and the members attending the Summer Institute 2002 held at 
Fuji-Yoshida and at the research meeting held in Nov.  for instructive discussions. 
M.~B.\  is supported in part by
the Grant-in-Aid for Scientific Research 
No.~12640295 from Japan Society for the Promotion of Science, and 
Grants-in-Aid for Scientific Purposes (A)  
``Neutrinos" (Y.~Suzuki) No.~12047225, 
from the Ministry of Education, Science, Sports and Culture, Japan.
%
%
%
%

\end{document}